\documentclass{article}

\usepackage{arxiv}

\usepackage[utf8]{inputenc} % allow utf-8 input
\usepackage[T1]{fontenc}    % use 8-bit T1 fonts
\usepackage{hyperref}       % hyperlinks
\usepackage{url}            % simple URL typesetting
\usepackage{booktabs}       % professional-quality tables
\usepackage{amsfonts}       % blackboard math symbols
\usepackage{nicefrac}       % compact symbols for 1/2, etc.
\usepackage{microtype}      % microtypography
\usepackage{lipsum}
\usepackage[dvips]{graphicx}

\graphicspath{ {./images/} }

\title{Phonon and exciton temperature-dependent properties of twisted MoS$_2$ }

\author{
 Francisco Silva-Santos \\
  Graduation Program of Physics\\
  Federal University of Piaui\\
  Teresina 64049-550, PI, Brazil \\
   \And
 Ruben da Silva \\
  Department of Physics\\
  Federal University of Piaui\\
  Teresina 64049-550, PI, Brazil \\
  \And
 Yuset Guerra Dávila \\
  Department of Physics\\
  Federal University of Piaui\\
  Teresina 64049-550, PI, Brazil \\
   \And
 Maykol Oliveira \\
  Department of Physics\\
  Federal University of Piaui\\
  Teresina 64049-550, PI, Brazil \\
   \And
 Zhuohang Yu\\
  Department of Physics and Center for 2-Dimensional and Layered Materials\\
  ThePennsylvania State University\\
  University Park, PA 16802, USA \\
  \And
 Mauricio Terrones\\
  Department of Physics and Center for 2-Dimensional and Layered Materials\\
  ThePennsylvania State University\\
  University Park, PA 16802, USA \\
   \And
 Antonio Souza Filho\\
  Department of Physics\\
  Federal University of Ceará\\
  Fortaleza, Ceará, 60455-900 Brazil \\
  \And
 Rafael Alencar\\
  Department of Physics\\
  Federal University of Ceará\\
  Fortaleza, Ceará, 60455-900 Brazil \\
  \texttt{rafael\_alencar@fisica.ufc.br} \\
   \And
 Bartolomeu Viana\\
  Department of Physics\\
  Federal University of Piaui\\
  Teresina 64049-550, PI, Brazil \\
  \texttt{bartolomeu@ufpi.edu.br} \\
}

\begin{document}
\maketitle
\begin{abstract}
In the present work, Raman and photoluminescence spectroscopies were used to study the dynamics of phonons and different excitons of MoS$_2$ bilayer under a rotation of 29$^{\circ}$ dependent of the temperature. The twisted bilayer (T-2L) of MoS$_2$ was obtained through mechanical exfoliation, and subsequently rotated using a dimethyl polysiloxane (PDMS) substrate and deterministic transferred to a SiO$_2$ substrate. The Raman spectrum of the twisted bilayer presents three peaks E$'$ (386 cm$^{-1}$), A$'_{1}$ (405 cm$^{-1}$) and an FA$'$ peak at approximately 409 cm$^{-1}$ linked to the A$'_{1}$ mode, which is attributed to a Moiré pattern phonon. Both modes (A$'_{1}$ and FA$'$) are dependent of light polarization in a way that demonstrates an effective coupling between the layers. It was also verified through the Grüneisen parameter, an increase in the anharmonicity of the mode in the E$'$ plane and a decrease the same for A$'_{1}$. In the PL measurements, the appearance of an exciton in T-2L was verified, which generated a second shoulder measured at $\approx$1.58 eV attributed to an indirect transition of an I trion. The interaction between the rotated monolayers of MoS$_{2}$ proved to be an important parameter for possible fine-tuning of the properties of bilayer samples.
\end{abstract}

% keywords can be removed
\keywords{MoS$_2$ \and monolayer \and twisted \and phonon \and trion}

\section{Introduction}
The studies involving layered and two-dimensional (2D) materials gained prominence with the experimental isolation of graphene. With this achievement, a range of other possibilities opens up for exploring a wide and rich variety of phenomena involving the electronic structures of other materials \cite{Rao2014}. 2D materials from the transition metal dichalcogenide (TMD) family, such as molybdenum disulfide (MoS2), tungsten diselenide (WSe2), tungsten disulfide (WS2), and molybdenum diselenide (MoSe2), due to quantum confinement effects, present optical and electronic properties that are interesting due to the possibilities of applications in a variety of devices such as electronics, photonics, energy storage, and optoelectronics \cite{Choi2017,Zhou2022}. An attractive aspect of these materials is the fact that their electronic and vibrational properties are strongly dependent on the number of layers. Molybdenum disulfide (MoS2) exhibits a strong photoluminescence signal that decreases dramatically when the system goes from its monolayer structure to its bulk form, which happens due to the transition process from direct to indirect bandgap \cite{Splendiani2010,Kadantsev2012}.  As well as other properties related to light absorption, photocurrent, and valleytronics \cite{Jiang2018,Lee2017,Mai2013} , which are also susceptible to layer coupling effects, showing the importance of the effects arising from the interactions in the stacking of 2D material layers. In particular, structures formed by MoS$_2$ bilayer demonstrate optical and electronic properties characteristic for the development of photoelectronic detectors \cite{Wang2021} and transistors \cite{Wang2012}. 
The relative rotation angle in the coupling between layers of 2D materials has attracted attention from researchers due to the possibility of promoting fine-tuning in the electronic structure of the material, inducing a strong electron correlation phenomenon for certain rotation angles due to the effects of the formation of the Moiré potential lattice \cite{Wu2020}. Therefore, electronic and optical properties demonstrate a dependence on the rotation angle \cite{Liao2020,Huang2016}. The Raman spectrum of MoS$_2$ exhibits characteristic peaks associated with the lattice modes in ultra-low-frequency (< 50 cm$^{-1}$), out-of-plane breathing modes, and in-plane shear mode, which are more susceptible to the rotation angle compared to the intralayer or high-frequency modes, E$_{1g}$ and A$_{1g}$  \cite{Huang2016}. Thus, Raman spectroscopy becomes very promising for identifying changes in the phonon dynamics in twisted MoS$_2$ bilayers. 
In this work, we investigated the behavior of the vibrational modes and PL emission properties of the MoS$_2$ monolayer and twisted bilayer with a relative twist angle of 30$^{\circ}$,  an angle at which the non-periodicity in the formation of the Moiré potential network is more evident and at which there has not yet been much study of these properties for this configuration of the MoS$_2$ twisted bilayer.  We used Raman spectroscopy and photoluminescence (PL) spectroscopy dependent on temperature and excitation power in order to analyze the correlation with thermodynamic parameters, specifically the relationship of the Grüneisen parameters for the isobaric mode ($\gamma_{iP}$). To gain a deeper understanding of the defect states or energy traps on the surface of MoS$_2$, we conducted a study on the temperature-dependent behavior of photoluminescence (PL). In addition to analyzing additional insights into the exciton and trion binding energies in monolayer and bilayer MoS$_2$.   

\section{Materials and Methods}
To prepare the twisted bilayer MoS$_2$ sample, we initiated the process by mechanically exfoliating a monolayer MoS$_2$ onto a polydimethylsiloxane (PDMS) stamp. Subsequently, half of this monolayer was deposited onto a SiO$_2$ substrate. The rotated configuration was achieved by subjecting the substrate to a 29$^{\circ}$ rotation (in relation to the sample on the PDMS stamp) and subsequently stamping the second portion of the monolayer MoS$_2$. This protocol was designed to ensure consistent crystallographic orientation prior to the rotation of both monolayers.
Raman and Photoluminescence (PL) micro-spectroscopies were performed on a LabRam Evolution spectrometer from Horiba equipped with confocal optical microscope Olympus BX41 and visible objective lens of 100X (NA = 0.9, WD = 0.21 mm) at room temperature and a CCD synapse as a detector. 1800 grooves/mm and 600 grooves/mm gratings were used to get the Raman and PL spectra, respectively. For low and high temperatures measurements a Linkam THMS 600 stage was used to control the temperature of the sample and a 50X LWD visible objective (NA = 0.50, WD = 10.6 mm) to focus the samples inside the cell. A solid-state laser with 532 nm wavelength was used to excite the sample and the power to reach it was controlled by a neutral density filter. The polarized Raman experiments were performed by rotating the linearly polarized incident laser beam through a half-wave plate and the scattered light was analyzed by a linear polarizer.

\section{Results and Discussion}
\begin{figure}[h]
\centering
\includegraphics[width=0.8\textwidth]{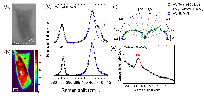}
\caption{(a) shows an optical image of the twisted 2L-MoS$_2$. The twist angle of the two monolayer MoS$_2$ is 29. Figure (b) is an E$'$ Raman map. The Raman spectrum of twisted bilayer MoS$_2$ (figure (c)) was acquired at the white dot in Figure (b), while the Raman spectrum from 1L was acquired at the black dot. The FA$^{'}_{1}$ is a folded A$^{'}_{1}$ vibrational mode which appears due to the folding phonon branches. Figure (d) shows that all modes, labeled as A$^{'}_{1}$, have an A$^{'}_{1}$ irreducible representation (totally symmetric mode). Figure (e) shows the low-frequency Raman mode in T-2L called breathing mode (BM))}
\label{fig:fig1}
\end{figure}

The regions corresponding to monolayer (1L) and twisted bilayer MoS$_2$ (2L), where Raman spectroscopy measurements were conducted, can be observed in Figure \ref{fig:fig1}. Figure \ref{fig:fig1}(a) shows an optical microscopy image of the sample, while Figure \ref{fig:fig1}(b) displays a Raman mapping image which indicate the distribution of Raman signal intensities associated with the E$'$ mode in both regions of the sample. The twisted bilayer MoS$_2$ is the reddish area, while the greenish areas are from 1L-MoS$_2$. In order to prepare the sample, we deposited monolayer MoS$_{2}$ on the PDMS substrate. After, we used the same monolayer to prepare the twisted bilayer MoS$_2$. This ensures the same crystallography orientation before the rotation of both monolayers. In Figure \ref{fig:fig1}(c), we observe the Raman spectra of both the monolayer region (lower spectrum) and the twisted bilayer region (upper spectrum) of MoS$_2$. These spectra were acquired at the black and white points in Figure \ref{fig:fig1}(b), respectively, at room temperature using 532 nm laser excitation. The Raman spectrum display peaks associated with the dominant Raman modes of MoS$_2$, related at the in-plane and out-of-plane vibrations and, though the analysis of group theory, these modes are understood as E$^{1}_{2g}$ and A$_{1g}$ at the center of the Brillouin zone, respectively, when describing the vibrations in the bulk form. However, by convention, when associated with the Raman spectrum of few layers, they are often denoted as E$'$ and A$^{i}_{1}$, reflecting their reduced symmetry compared to the three-dimensional crystal. Furthermore, the relative difference in the position between these two modes is employed as one key parameters for the rapid identification of the number of MoS$_2$ layers, as reported in Raman spectroscopy studies on MoS$_2$ layered structures \cite{Li2012,Lee2010,Niu2018}. In particular, the Raman spectrum of twisted bilayer exhibits a peak at around 409 cm$^{-1}$, which correlates with the degree of rotation between the layers. This Raman mode is attributed to a Moiré phonon that is bound to a segment of the A$^{'}_{1}$ phonon, referred to as FA$^{'}_{1}$. The position of FA$^{'}_{1}$ reveals a sinusoidal behavior in response to changes in the rotation angle, assuming a position around 410 cm$^{-1}$ for an angle of $\sim$ 30$^{\circ}$ \cite{Liao2020,Lin2018}. This behavior arises from the differing wave vectors in the phonon dispersion. In addition, the intensity of the FA$^{'}_{1}$ mode experiences an exponential increase when the rotation angle varies from 8$^{\circ}$ to 30$^{\circ}$,  due to a sharp increase in the density of Moiré phonons \cite{Liao2020}. Such parameters related to the rotation angle confirm our Raman experiment data regarding the angle of 29$^{\circ}$ in the twisted bilayer of the MoS$_2$ sample analyzed in this study. The intensity of the A$^{'}_{1}$ peak in both the 1L and 2L regions, as well as the FA$^{'}_{1}$ peak in the twisted bilayer, displays a dependence with the polarization of the incident light (relative to the orientation of the polarizer). The peaks are at their maximum at 0$^{\circ}$ (180$^{\circ}$) and reach their minimum at 90$^{\circ}$ (270$^{\circ}$), as illustrated in the polar plot in Figure \ref{fig:fig1}(d). These peaks exhibit the expected behavior for a mode that possesses complete symmetry within the MoS$_2$ structure. An important vibration mode in the analysis of mechanical and structural properties of two-dimensional materials, which highlights an effective coupling between the layers \cite{Lin2018},  is known as the "breathing mode" (BM). This mode involves the synchronized expansion and contraction of atomic layers, rhythmically opening and closing. Figure \ref{fig:fig1}(e) illustrates the BM in a twisted bilayer of MoS$_2$.

\begin{figure}[h]
\centering
\includegraphics[width=0.8\textwidth]{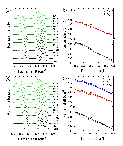}
\caption{(a) and (c) Raman spectra of MoS$_2$ monolayer and bilayer region, respectively, recorded between 93 and 573 K showing the phonon dependence as temperature. (b) and (d) shows the temperature dependence of the Raman frequency shift of E$^{1}_{2g}$ and A$_{1g}$ modes for monolayer region and for bilayer with the FA$_{1g}$ (written as R-A$_{1g}$ in Figure (d)) mode appearing. }
\label{fig:fig2}
\end{figure}

\begin{figure}[h]
\centering
\includegraphics[width=0.98\textwidth]{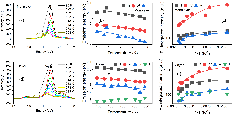}
\caption{Photoluminescence (PL) spectra of 1L (a) and T-2L (d) recorded at each temperature value between 93 K and 293 K reveal the temperature-dependent emission bands. Figures (b) and (e) highlight the intensity variation with temperature (data points) for exciton A$^0$, trion A$^-$, and exciton B in monolayer and twisted bilayer MoS$_{2}$, respectively, along with the best fits (lines) according to equation (\ref{eq:eq1}.  Figures (c) and (f) show the PL intensity (data points) as a function of 1/T for each emission band, with adjustments (lines) based on equation \ref{eq:eq2}. The behavior of the exciton associated with the indirect transition I is also presented.}
\label{fig:fig3}
\end{figure}

Figure \ref{fig:fig2}(a) and \ref{fig:fig2}(c) displays the Raman spectra of the monolayer and bilayer region, respectively, for each temperature. Temperature-dependent Raman measurements were performed range spanning from 93 to 573 K using a 532 nm laser as excitation source, illustrating how phonon behavior changes with temperature. The Raman spectrum of the MoS$_{2}$ mono and bilayer for the high frequency spectral region (370 to 420 cm$^{-1}$) is consistent with the two active Raman modes E$^{1}_{2g}$ and A$_{1g}$.  According to Suman Sarkar et al., due to the antiphasic oscillations in and out of the plane of the S atoms, the E and A modes arise respectively.  Only the vibrations of the S atoms contribute to the A mode, to the E mode they arise from the two atoms Mo-related S. Shifts towards red are observed in the E$^{1}_{2g}$ and A$_{1g}$ modes, when the temperature varies between 93 – 293 K and between 303 – 573 K.  As already observed in other layered materials,  the wavenumber of E$^{1}_{2g}$ and A$_{1g}$ modes decrease with increasing temperature \cite{Liao2020,Lin2018}. To adjust the Raman spectra at each temperature, two and three vibrational peaks were used for mono and bilayer, Figure \ref{fig:fig3}(b) and \ref{fig:fig3}(d), respectively. The solid curves are adjusted to the data obtained with a quadratic polynomial fit,  following the equation \ref{eq:eq1}: 

\begin{equation}\label{eq:eq1}
\omega(T)=\omega_{0}+X_{1}T+X_{2}T^2  
\end{equation}

is used to best fit the nonlinear temperature-dependence Raman shift data, where $\omega_0$ is Raman shift at zero Kelvin \cite{Yang2016}. The fitting gives out the first-order and second-order temperature coefficients, X$_1$ (cm$^{-1}/K$) and X$_2$ (cm$^{-1}/K{^2}$), respectively, but only X$_1$ was listed in Table \ref{tab:table1}. The primary reason for the alteration in the Raman shifts due to temperature effects predominantly arises from the influence of anharmonic effects involving three- and four-phonon processes that can be describe though anharmonic phonon decay model, based on the Balkanski approach \cite{Balkanski1983}:

\begin{equation}\label{eq:eq2}
\omega(T)=\omega_{0}+A\left[1+\frac{2}{e^{x}-1}\right]+B \left[1+\frac{3}{e^{y}-1}+\frac{3}{(e^{y}-1)^{2}}\right]
\end{equation}

where $x= \hbar \omega_{0} /2K_{B}T$, $y= \hbar\omega_{0} /3K_{B}T$, T is the absolute temperature, $\omega(T)$ is the wave number at T, k$_{B}$ is the Boltzmann constant. Parameters A and B are constants and are related to the contributions of three- and four-phonon processes to the frequency changes. In the high-temperature limit, the factors multiplying A and B in Eq. (\ref{eq:eq2}) vary as T and T$^{2}$, respectively.

The Grüneisen parameter $\gamma$ plays a crucial role in thermodynamics by quantifying how variations in temperature and pressure affect the volume of a cell or material.  It serves as a valuable metric for assessing the degree of anharmonicity in the phonon spectrum, with higher values of $\gamma$ indicating greater levels of anharmonicity. When considering the relationship of the Grüneisen parameters for the isobaric mode ($\gamma_{iP}$),  we employ the following equation \cite{Yang2017}:

\begin{equation}\label{eq:eq3}
    \gamma_{iP}=- \left( \frac{\partial \ln \omega_{0}}{\partial \ln V}\right)_{P}=-\frac{1}{\alpha \omega_{0}}\left( \frac{\partial \omega}{\partial T}\right)_{P}
\end{equation}

where $\alpha$ parameter is the thermal expansion coefficient, $\omega_{0}$ and $\partial \omega/\partial T=X_{1}$ are defined in Eq (\ref{eq:eq1}), estimated from the experimental data and listed in Table \ref{tab:table1} for each mode Raman. 

\begin{table}
 \caption{E$^1_{2g}$, A$_{1g}$ and FA$_{1g}$ mode Raman shift at T = 0 K ($\omega_0$), the first-order temperature coefficient  X$_1$ ($\partial \omega / \partial T$) and the Grüneisen parameter for isobaric mode ($\gamma_{iP}$) of monolayer (1L) and twisted bilayer MoS2 (T-2L).}
  \centering
  \begin{tabular}{ccccc}
   \toprule
    %\multicolumn{2}{c}{Part}                   \\
    %\cmidrule(r){1-2}
   MoS$_2$     & Vibrational mode     &$\omega_0$ (cm$^{-1}$) &X$_{1}$ (cm$^{-1}/K$)&$\gamma_{iP}$ \\
    \midrule
     % \toprule
    \multicolumn{1}{c}{1L}                   \\
   % \cmidrule(r){1}
     & E$^{1}_{2g}$  & 390.23$\pm$ 0.26&-0.0130$\pm$0.0017&1.92$\pm$ 0.25     \\
     & A$_{1g}$ &408.30$\pm$0.37&-0.00872$\pm$0.0024&1.23 $\pm$0.33    \\
    \midrule
        \multicolumn{1}{c}{T-2L}                   \\
   % \cmidrule(r){1}
     & E$^{1}_{2g}$  & 389.17$\pm$0.25&-0.0120$\pm$0.0017&2.47$\pm$ 0.35   \\
     & A$_{1g}$ &407.58 $\pm$0.35&-0.00466$\pm$0.0023&0.92 $\pm$ 0.45    \\
     & FA$_{1g}$  & 411.68 $\pm$0.26&-0.00362$\pm$ 0.0017&0.71 $\pm$0.33    \\
     & BM&36.25$\pm$0.30&-0.00244$\pm$ 0.0021&5.41 $\pm$ 4.65   \\
    \bottomrule
  \end{tabular}
  \label{tab:table1}
\end{table}

In order to calculate $\gamma_{iP}$, we used the value of the thermal expansion coefficient ($\alpha$) reported for the monolayer of MoS$_2$ at room temperature, which is defined as 1.74$\cdot$10$^{-5}$ K$^{-1}$ \cite{Yang2017,Bhatt2014}.  This same value was used in this work to evaluate the effects of anharmonicity on the phonons observed in the MoS$_2$ monolayer. On the other hand, for the bilayer, we used the coefficient of thermal expansion ($\alpha$) that has been previously reported in other works as the value for bulk MoS$_2$, defined as 1.245$\cdot$10$^{-5}$ K$^{-1}$ \cite{Yang2017,Peng2016}. Table \ref{tab:table1} displays the calculated values of $\gamma_{iP}$ for Raman modes in the monolayer (1L) and twisted bilayer (T-2L) of MoS2, along with their respective uncertainties. The results indicate that the degree of anharmonicity in the E$^{1}_{2g}$ mode increases from 1L to T-2L, while the anharmonicity of the A$_{1g}$ mode decreases with an increasing number of layers. For the out-of-plane acoustic mode (FA$_{1g}$), present only in the bilayer,  the calculated value of $\gamma_{iP}$ is approximately 0.71$\pm$ 0.33.

At room temperature, the PL spectrum of 1L shows a broad, high-intensity, asymmetric band around $\sim$1.85 eV and a low-intensity shoulder at $\sim$2 eV. The high-intensity band was adjusted with two Lorentzian peaks, associated with the A$^{0}$ excitation and the negatively charged A- trion, which is a bound state, while the shoulder associated with the B excitation was adjusted with one peak, see Figure \ref{fig:fig3}. In the PL spectrum of 2L, which also shows a broad band at around $\sim$1.85eV, two low-intensity shoulders are observed: one associated with trion I at around 1.6 eV and the other associated with excitation B at $\sim$2 eV, also adjusted with the Lorentzian. To gather more insights into the defect states or surface energy traps of MoS$_2$, we conducted a study on the temperature-dependent photoluminescence (PL) emission of both monolayer (1L) and twisted bilayer (T-2L) at low temperatures ranging from 93 K to 293 K  \cite{Christopher2017}.  As depicted in Figure \ref{fig:fig3},  as the temperature increases from 93 K to room temperature, the lines positioned at $\sim$1.82, $\sim$1.83, and $\sim$2.01 eV undergo a redshift, indicative of thermal effects. Meanwhile, the line at $\sim$1.58 eV, observed exclusively in T-2L, experiences a blueshift. As anticipated, the PL intensity diminishes with rising temperature \cite{Ranjuna2023,Golovynskyi2020}. These temperature-dependent energy shifts are elucidated using the model proposed by O'Donnell et al \cite{ODonnell1991}:

\begin{equation}
E_{g}(T)=E_{g}(0)-S\langle \hbar \omega\rangle\left[\coth \left(\frac{\langle\hbar \omega\rangle}{2k_{B}T}\right)-1\right]
\end{equation}

Here, E$_g$(0) represents the band gap at 0 K, S is the coupling constant, $\langle \hbar \omega \rangle$ is the average electron-phonon interaction energy, and k$_B$ is the Boltzmann constant. The E$_g$(0) values are 1.89/2.03 eV for A$^0$/B excitons in 1L and T-2L flakes, while for the A$^-$ trion, the values are 1.83 for 1L and 1.81 for T-2L.  The $\langle \hbar \omega \rangle$ values are 34/53 for A$^0$/B and 2.45/4.83 for S. The A$^-$ trion in 1L has a $\langle \hbar \omega \rangle$ value of 39 eV.  Refer to Table \ref{tab:table2} for $\langle \hbar \omega \rangle$ and S values for T-2L. 

\begin{table}
 \caption{Parameters describing the A$^0$ and B excitons and the A$^-$ trion for 1L and 2L MoS$_2$.}
  \centering
  \begin{tabular}{ccccc}
   \toprule
    %\multicolumn{2}{c}{Part}                   \\
    %\cmidrule(r){1-2}
   MoS$_2$     &Exciton &E$_g$(0) (eV) &S&
   $\langle \hbar \omega \rangle$ (meV) \\
    \midrule
     % \toprule
    \multicolumn{1}{c}{1L}                   \\
   % \cmidrule(r){1}
     &  B & 2.03&4.83&	53  \\
     &  A$^{0}$&1.89&2.45&34    \\
     &  A$^{-}$&1.83&	5.03	&39    \\
    \midrule
        \multicolumn{1}{c}{T-2L}                   \\
   % \cmidrule(r){1}
     &  B & 2.03&	1.91	&25.4 \\
     &  A$^{0}$&1.89	&1.81	&22   \\
     &  A$^{-}$&1.81&	2.93	&34    \\
    \bottomrule
  \end{tabular}
  \label{tab:table2}
\end{table}

To gather additional insights into the binding energies of excitons and trion, we fitted the three emission peaks of 1L and the four peaks of T-2L using an integrated Lorentz function. We extracted the photoluminescence (PL) intensities of these peaks and plotted them as a function of 1/T. As the temperature increases, the intensities of these peaks gradually increase and then tend to decrease. The results were analyzed using the multilevel model provided by \cite{Lin2012,Huang2016a,Shibata1998}:

\begin{equation}
I(T)=I(0) \frac{1+Ae^{-E_{1}/k_{B}T}}{1+Be^{-E_{2}/k_{B}T}}
\end{equation}

where the PL intensity at 0 K is given by I(0), the Boltzmann constant is k$_B$, A and B are fitting parameters. The activation energies are described by E$_1$ and E$_2$, the values obtained for 1L and T-2L are shown in Table \ref{tab:table3}.

\begin{figure}[h]
\centering
\includegraphics[width=0.8\textwidth]{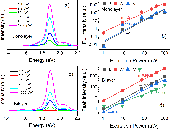}
\caption{The photoluminescence spectra of the monolayer (1L) and the twisted bilayer (T-2L) in relation to the excitation power are shown in (a) and (b), ranging from 1.1 to 1200 $\mu$W. Figures (c) and (d) illustrate the linear behavior of the emission intensity in relation to the power.}
\label{fig:fig4}
\end{figure}

Figure \ref{fig:fig4} shows the PL spectra of 1L and T-2L at six different excitation powers. It can be seen that all the bands observed in the PL with varying temperature are observed with varying excitation power.  As the laser excitation power increases, an increase in the intensity of the A$^0$ band is observed in both 1L and T-2L. It is important to note that practically all the bands show no shift with increasing laser power excitation, neither for red nor for blue,  and only exhibit an almost linear dependence on excitation intensity.

\begin{table}
 \caption{Parameters describing the A$^0$ and B excitons and the A$^-$ trion for 1L and 2L MoS$_2$.}
  \centering
  \begin{tabular}{cccc}
   \toprule
    %\multicolumn{2}{c}{Part}                   \\
    %\cmidrule(r){1-2}
   MoS$_2$     &Exciton &E$_1$ (meV) &E$_2$ (meV) \\
    \midrule
     % \toprule
    \multicolumn{1}{c}{1L}                   \\
   % \cmidrule(r){1}
     &  B & 193	&23 \\
     &  A$^{0}$&31	&52   \\
     &  A$^{-}$&0.16	&101   \\
    \midrule
        \multicolumn{1}{c}{T-2L}                   \\
   % \cmidrule(r){1}
     &  B & 4.4	&53.9 \\
     &  A$^{0}$&25.7	&18.3   \\
     &  A$^{-}$&115 &	8.9 \\
      &  I&33 &	48 \\
    \bottomrule
  \end{tabular}
  \label{tab:table3}
\end{table}

\section{Conclusions}

The study on MoS$_2$, particularly focusing on its twisted structure at 29 degrees, unfolded essential insights into various vibrational and electronic aspects. The investigation delved into the intricate phonon and exciton dynamics of MoS$_2$, shedding light on key features such as the FA’ and the breathing modes. Moreover, the exploration of the first-order temperature coefficient of Raman modes in both monolayer and twisted bilayer MoS$_2$ offered a comprehensive understanding of the material's response to temperature variations. A significant aspect of the research involved analyzing the isobaric mode Grüneisen parameter in T-2L-MoS$_2$. This parameter serves as a crucial indicator of how the material's vibrational properties respond to external pressure changes. The findings contribute to our understanding of MoS2's behavior under different environmental conditions. It was observed an increasing of the anarmonicity for the E$^{1}_{2g}$ and and decreasing for A$_{1g}$. The anarmonicity of the FA$_{1g}$ and BM modes were estimated, as well, for the first time for T-2L-MoS$_2$. Furthermore, the study extended its focus to the exciton parameters, exploring their temperature dependence in both monolayer and twisted bilayer MoS$_2$. 
Excitons play a pivotal role in the material's optical and electronic properties, and understanding how these parameters vary with temperature provides valuable insights into the material's potential applications in different thermal environments. The electron-phonon interaction energies were estimated and a decreasing of this interaction is noticed for excitons in T-2L-MoS$_2$. A notable highlight of the research involves the determination of activation energies for specific excitons (A$^0$, B excitons, and trion A$^-$). These activation energies provide crucial information about the energy barriers associated with excitonic processes, offering a deeper understanding of the material's electronic structure and potential applications in optoelectronics. The activation energies showed a decreasing for excitons in T-2L-MoS$_2$ and is inverted for the trion A$^-$. The excitation energies for the I exciton was estimated for T-2L-MoS$_2$. In conclusion, the comprehensive investigation into MoS$_2$, considering its twisted structure, vibrational modes, temperature-dependent exciton parameters, and activation energies, significantly advances our knowledge of the material's fundamental properties. These findings not only contribute to the broader understanding of MoS$_2$ but also pave the way for potential applications in electronic and optoelectronic devices with enhanced control over its structural and vibrational characteristics promoting a engineering of the electronic properties by the angle of rotation between the layers.

\bibliographystyle{unsrt}  


%\begin{thebibliography}{1}
%
%\bibitem{kour2014real}
%George Kour and Raid Saabne.
%\newblock Real-time segmentation of on-line handwritten arabic script.
%\newblock In {\em Frontiers in Handwriting Recognition (ICFHR), 2014 14th
%  International Conference on}, pages 417--422. IEEE, 2014.
%
%\bibitem{kour2014fast}
%George Kour and Raid Saabne.
%\newblock Fast classification of handwritten on-line arabic characters.
%\newblock In {\em Soft Computing and Pattern Recognition (SoCPaR), 2014 6th
%  International Conference of}, pages 312--318. IEEE, 2014.
%
%\bibitem{hadash2018estimate}
%Guy Hadash, Einat Kermany, Boaz Carmeli, Ofer Lavi, George Kour, and Alon
%  Jacovi.
%\newblock Estimate and replace: A novel approach to integrating deep neural
%  networks with existing applications.
%\newblock {\em arXiv preprint arXiv:1804.09028}, 2018.
%
%\end{thebibliography}

\end{document}